# Evidence for Band Renormalizations in Strong-coupling Superconducting Alkali-fulleride Films

J. S. Zhou[1,2], R. Z. Xu[1,2], X. Q. Yu[1,2], F. J. Cheng[1,2], W. X. Zhao[1,2], X. Du[1,2], S. Z. Wang[1,2], Q. Q. Zhang[1,2], X. Gu[1,2], S. M. He[3], Y. D. Li[1,2], M. Q. Ren[1,2], X. C. Ma[1,2,6], Q. K. Xue[1,2], Y. L. Chen[3,4,5*], C. L. Song[1,2,6], and L. X. Yang[1,2,6*]

[1]*State Key Laboratory of Low Dimensional Quantum Physics, Department of Physics, Tsinghua University, Beijing 100084, China.*
[2]*Frontier Science Center for Quantum Information, Beijing 100084, China.*
[3]*Department of Physics, Clarendon Laboratory, University of Oxford, Parks Road, Oxford OX1 3PU, UK.*
[4]*School of Physical Science and Technology, ShanghaiTech University and CAS-Shanghai Science Research Center, Shanghai 201210, China.*
[5]*ShanghaiTech Laboratory for Topological Physics, Shanghai 200031, China.*
[6]*Collaborative Innovation Center of Quantum Matter, Beijing 100084, China*

**There has been a long-standing debate about the mechanism of the unusual superconductivity in alkali-intercalated fulleride superconductors. In this work, using high-resolution angle-resolved photoemission spectroscopy, we systematically investigate the electronic structures of superconducting $K_3C_{60}$ thin films. We observe a dispersive energy band crossing the Fermi level with the occupied bandwidth of about 130 meV. The measured band structure shows prominent quasiparticle kinks and a replica band involving high-energy Jahn-Teller active $H_g(8)$ phonon mode, reflecting strong electron-phonon coupling in the system. The electron-phonon coupling constant is estimated to be about 1.2, which dominates the quasiparticle mass renormalization. Moreover, we observe an isotropic nodeless superconducting gap beyond the mean-field estimation ($2\Delta/k_B T_c \approx 5$). Both the large electron-phonon coupling constant and large reduced superconducting gap suggest a strong-coupling superconductivity in $K_3C_{60}$, while the electronic correlation effect is suggested by the observation of a waterfall-like band dispersion and the small bandwidth compared with the effective Coulomb interaction. Our results not only directly visualize the crucial band structure of superconducting fulleride but also provide important insights into the mechanism of the unusual superconductivity.**

Alkali-intercalated fullerides $A_3C_{60}$ ($A$ = K, Rb, Cs) not only record the superconducting transition temperature among the molecular superconductors [1-7] but also exhibit many unusual properties that resemble cuprate and iron-based high-temperature superconductors, such as domed superconducting phase diagram [4], the proximity to a magnetic Mott-insulating parent state [8-13], and the formation of a pseudogap [14]. After extensive research efforts, the mechanism of the superconductivity in fullerides, however, remains controversial due to the strong entanglement of electronic correlation effect and complicated electron-phonon coupling (EPC). While fulleride superconductors were arguably considered to be conventional Bardeen-Cooper-Schrieffer (BCS) superconductors [15-18], many unconventional superconducting mechanisms have been proposed [11-13,19-21], including polaron-driven superconductivity [22,23], local pairing with Jahn-Teller phonons assisted by Coulomb repulsion [13,24], negative Hund's coupling stabilized by EPC [25], and pure electronic pairing [26]. The interplay between microscopic interactions underlying the electronic phase diagram of fullerides also remains elusive up to date.

To understand the unusual superconductivity in fullerides, it is highly desired to investigate their electronic structure in the energy-momentum space. From the electronic structure perspective, the lowest unoccupied molecular orbital (LUMO) of $C_{60}$ molecule is of $t_{1u}$ symmetry [17]. Jahn-Teller distortion with possible $D_{2h}$ symmetry further lifts the degeneracy of the $t_{1u}$ bands and the three electrons donated by alkali atoms populate the split $b_{2u}$ and $b_{3u}$ bands of $A_3C_{60}$, inducing a low-spin $S = ½$ ground state [11]. The conduction bandwidth $W$ is shown to be much smaller than the effective Coulomb interaction $U$ [27,28], and a Mott-Jahn-Teller insulator phase is discovered in $Cs_3C_{60}$ [8-12]. These observations suggest an important role of the electronic correlation in the electronic properties of $A_3C_{60}$ [11-13], although no Mott localization is observed in superconducting $K_3C_{60}$ and $Rb_3C_{60}$. Despite these important experimental and theoretical breakthroughs, direct measurement of the electronic band structure and its coupling to phonon modes is still essentially lacking due to the lack of high-quality crystal surfaces. In particular, the momentum distribution of the superconducting gap is also yet to be experimentally investigated.

In this work, we overcome the obstacle of the sample-surface quality by synthesizing high-quality ultrathin films of K-intercalated $C_{60}$ on bilayer graphene and systematically investigate their electronic structures using high-resolution angle-resolved photoemission spectroscopy (ARPES). We present the evolution of the electronic structure with film thickness, K-doping, and temperature. While the electronic structure of slightly K-doped $C_{60}$ shows a large insulating gap, $K_3C_{60}$ exhibits a highly dispersive band crossing the Fermi level ($E_F$), which is strongly renormalized by multiple phonon modes, directly evidencing the strong EPC in the system. Interestingly, we observe a replica band due to the formation of polarons involving Jahn-Teller active $H_g(8)$ phonon at about 192 meV. In the superconducting state, we observe an isotropic nodeless superconducting gap, suggesting an s-wave superconducting pairing. The magnitude of the reduced superconducting gap $2\Delta/k_BT_c \approx 5$ is beyond mean-field estimation, which, together with a large EPC constant of 1.2, suggests a strong-coupling superconductivity in $K_3C_{60}$. In addition, we observe waterfall-like band dispersion in a large energy scale, alluding the importance of electronic correlation in the system, which is supported by the small ratio between the bandwidth $W$ and effective Coulomb interaction $U$. Our results not only directly visualize the long-sought crucial electronic structure of superconducting fullerides but also evidence the strong EPC involving multiple crucial phonon modes, which shed new light on the understanding of the unusual superconductivity of fullerides.

High-quality $K_3C_{60}$ films were prepared on bilayer graphene that was epitaxially grown on silicon carbide using molecular beam epitaxy (MBE) [Fig.1(a)] [14]. $C_{60}$ films were first synthesized layer-by-layer as monitored by the intensity oscillation of specular spot in the reflection high-energy electron diffraction (RHEED) pattern [Fig. 1(b)]. The K atoms were subsequently deposited on the $C_{60}$ films, followed by a slight annealing at room temperature for an hour so that K atoms can uniformly intercalate the $C_{60}$ films. The films were then *in-situ* transferred to ARPES chamber under ultrahigh vacuum below $1.5 \times 10^{-10}$ mbar. High-resolution ARPES measurements were conducted using a DA30 analyser and Scienta VUV5050 Helium lamp. The energy and angular resolutions were set to 7 meV and 0.2 ° respectively.

As schematically shown in Fig. 1(a), $K_3C_{60}$ films crystallize along the [111] direction of a face-centered cubic (fcc) structure. K atoms occupy either the tetrahedral (green spheres) or octahedral (red spheres) interstitial holes between the hexagonal $C_{60}$ layers. Figure 1(c) shows the surface topography of a 5 monolayer (ML) $K_3C_{60}$ film measured by scanning tunneling microscopy (STM). $C_{60}$ molecules show a uniform configuration with a hexagon facing up as manifested by the tri-lobe-like pattern. We observe no noticeable reconstruction and orientational disorder. The intermolecular distance is about $10.0 \pm 0.1$ Å, in good agreement with the previous results and the value in bulk fcc $K_3C_{60}$ [14,17,28]. Figure 1(d) shows the scanning tunneling spectroscopy (STS) measured along the dashed line in Fig. 1(c) at 4.7 K, in which the homogenous superconducting gap is clearly observed.

Figure 2 shows the evolution of the band structure of 3 ML $C_{60}$ films with K doping along $\bar{\Gamma}\bar{M}$. Without $C_{60}$, the epitaxial bilayer graphene substrate shows a gapped band structure around the $\bar{\Gamma}$ point with the band top at about 2.7 eV below $E_F$ (Supplemental Material, Fig. S1 [29]) [30]. The deposition of $C_{60}$ masks the band dispersion of the substrate and contributes dispersive bands around -2.5 eV and -3.9 eV, which are derived from the highest occupied molecular orbital (HOMO) and second HOMO (HOMO-1) of $C_{60}$ (Supplemental Material, Fig. S1 [29]) [31,32].

With K intercalation, the HOMO and HOMO-1 bands shift towards $E_F$ and become more dispersive while the band gap between them shrinks [Figs. 2(b-e)], which alludes an enhanced inter-molecular interaction that may be bridged by alkali metals. Prominently, extra electronic states emerge near $E_F$ in $K_3C_{60}$ due to the population of the LUMO band of $C_{60}$. These states dominate the density of states (DOS) of $K_3C_{60}$ near $E_F$ and are thus crucial for the superconductivity. The states near $E_F$ in $K_2C_{60}$, however, are due to the phase separation at slight K doping level (Supplemental Material,

Fig. S7 [29]). With further increasing K doping, the spectral weight of the newly emerged states is enhanced but shifts away from $E_F$ [Figs. 2(d-e)], suggesting a metal-to-insulator transition at doping levels much higher than x = 3 [28].

To understand the superconducting properties of $K_3C_{60}$, we investigate the temperature evolution of the fine band structure near $E_F$ in Figs. 3(a-e). We reveal highly dispersive hole band crossing $E_F$ at Fermi momentum $k_F = \pm 0.20 \pm 0.01$ Å$^{-1}$. This band contributes to the DOS peak near $E_F$ [Fig. 2(f)] and is therefore crucial for the novel superconductivity of $K_3C_{60}$. The occupied bandwidth $W$ is about 130 meV. Based on the band dispersion and the calculation, we estimate the full bandwidth to be less than 200 meV [27], much smaller than the effective Coulomb interaction $U$ of about 1 eV [28]. The Fermi velocity $v_F$ is estimated to be about $7.5 \pm 0.5 \times 10^6$ cm/s compared to $1.8 \times 10^7$ cm/s in the density-functional theory (DFT) calculation [33], suggesting an effective electron mass $m^* = (1 + \lambda_{ep} + \lambda_{ee}) m_b = (2.4 \pm 0.2) m_b$, where $\lambda_{ep}$ and $\lambda_{ee}$ are dimensionless EPC and electron-electron interaction parameters, and $m_b$ is the bare mass [34]. The area of the FS (Supplemental Material, Fig. S2 [29]) suggests an electron density of about $2.9 \pm 0.1$ e$^-$/unit cell [27], or $(3.3 \pm 0.1) \times 10^{14}$ e$^-$/cm$^2$ [$(4.1 \pm 0.1) \times 10^{21}$ e$^-$/cm$^3$]. Using the DOS of 7.2 eV$^{-1}$ per spin at $E_F$ [17,35], we obtain a small Fermi energy of about 0.3 eV supposing all the $t_{1u}$ electrons are free, consistent with previous results [36,37].

With decreasing temperature, the band dispersion becomes sharper while $k_F$ remains nearly unchanged [Figs. 3(a-d)]. At 8.5 K, we observe a strong renormalization of the band dispersion near -18, -54, and -85 meV, as shown by the dips in the integrated EDC in Fig. 3(e), which will be discussed later. The leading edge of the EDC at $k_F$ shifts towards high binding energies with decreasing temperature [Fig. 3(f)], evidencing the formation of superconducting gap, which is better visualized by the symmetrized ARPES spectra at 8.5 K in Fig. 3(g). By contrast, no gap is observed in the symmetrized spectra at 50 K [Fig. 3(h)]. Figure 3(i) shows the temperature evolution of the symmetrized EDCs at $k_F$. At 8.5 K, we observe a clear superconducting peak and superconducting gap [14]. With increasing temperature, the superconducting peak and superconducting gap gradually disappear. Interestingly, the symmetrized spectrum shows a dip at 30 K, which is an indication of a pseudogap that persists up to 40 K [14]. Figure 3(j) shows the temperature evolution of the leading-edge gap together with the temperature evolution of the gap depth extracted from scanning tunneling spectroscopy measurements (Supplemental Material, Fig. S4 [29]). The fit of the gap depth to the BCS-type temperature evolution function suggests a superconducting transition temperature of 20.6 K. The fit of the symmetrized EDC to the Dynes model gives a gap of about 4.3 meV at 8.5 K [black line in Fig. 3(i)], corresponding to the zero-temperature gap of about 4.5 meV. The reduced gap $2\Delta/k_B T_C$ is about 5, beyond the mean-field expectation of 3.5 [38]. Figure 3(k) presents the superconducting gap distribution in the momentum space (Supplemental Material, Fig. S3 [29]). Along different momentum directions, the superconducting gap remains constant within the accuracy of our experiment, which suggests a nodeless s-wave-like pairing symmetry. Consistently, the STS spectra show a U-shape superconducting gap at 4.7 K [Fig. 1(d)] [14].

Prominently, the band dispersion in Fig. 3(d) is strongly renormalized by multiple phonon modes [blue arrows in Figs. 3(d) and 3(e)]. We analyze the band renormalization by fitting the momentum distribution curves (MDCs) to Lorentzians (Supplemental Material, Fig.S5 [29]). The extracted band dispersions are shown in Fig. 4(a), from which we can observe the anomalies in the band dispersion near -54 meV and -85meV, without noticeable change with temperature. Based on a linear bare band assumption, we extract the real and imaginary parts of the electron self-energy in Figs. 4(b) and 4(c). We observe peak-like features and the change of the slope in the real and imaginary parts of electron self-energy respectively, which confirm the effect of EPC near -54 meV and -85 meV [thick gray lines].

The band renormalizations near -18 meV [manifested by the peak-dip-hump structure in Figs. 3(e) and 3(i)], -54 meV, and -85 meV can be attributed to the intermolecular phonon mode, intramolecular $H_g(2)$ phonon, and intramolecular $H_g(3)$ phonon, respectively [17,39]. According to previous experiments and calculations, the $H_g(2)$ phonon mode indeed couples stronger with the electrons than the $H_g(3)$ phonon [17,39]. Therefore, it induces a more noticeable kink in the band dispersion [Fig. 3(d)]. The coupling to the other two modes, on the other hand, are manifested by the peak-dip-hump structure and the reduction of the spectral weight [Fig. 3(e)].

Interestingly, we observe a replica band at about 192 meV below the main band as shown in Figs. 4(d) and 4(e). Due to the strong EPC in $K_3C_{60}$, we attribute this band replication to the formation of polaron involving the $H_g(8)$ phonon mode that has been shown to strongly couple to the electrons among the $H_g$ phonons in $K_3C_{60}$ [17,40,41]. Since this phonon has an energy larger than the bandwidth near $E_F$, it induces a replica band instead of a kink in the dispersion, similar to the observation of the replica band in FeSe/SrTiO$_3$ [42].

The EPC parameter can be derived from the temperature dependence of the imaginary part of electron self-energy at high temperatures by $\mathrm{Im}\Sigma(E_F, T) = \lambda_{ep}\pi k_B T$ [43]. As shown in Fig. S6 in the Supplemental Material [29], $\lambda_{ep}$ is estimated to be about $1.2 \pm 0.3$, about twice the value in previous results [15,17,40,44,45], suggesting a dominant role of EPC in the renormalization of effective electron mass. In principle, the superconducting transition temperature can be estimated by McMillan-Allen-Dynes formula: $T_c = \frac{\omega_{\ln}}{1.2}\exp(-\frac{1.04(1+\lambda_{ep})}{\lambda_{ep}-\mu^*(1+0.62\lambda_{ep})})$[46], where $\omega_{\ln}$ and $\mu^*$ are the logarithmic average phonon frequency and effective Coulomb pseudopotential respectively. Using $\omega_{\ln} \approx 50 \pm 10$ meV and $\mu^* \approx 0.25$ [47-50], $T_c$ is estimated to be $24 \pm 5$ K, consistent with the experimental value. It is noteworthy, however, both $\omega_{\ln}$ and $\mu^*$ vary with a large uncertainty in the literature [17,39,51]. With the phonon energy of about 100 meV, $\mu^*$ up to 0.35 is required to reproduce the experimental $T_c$ with McMillan equation [52]. Nonetheless, the revealed large EPC parameter, together with the large reduced superconducting gap $\frac{2\Delta}{k_B T_c} \approx 5$, suggests the strong-coupling nature of the superconductivity of $K_3C_{60}$.

On the other hand, we observe waterfall-like band dispersion in an energy window as large as 1.2 eV as shown in Fig. 4(d), reminiscent of the band dispersion of cuprate superconductors and other strongly correlated materials [53-55]. It suggests the impact of electronic correlation in the electronic structure, which is supported by the small *W*/*U* ratio [27,28], the small Fermi energy [56], and possibly large Coulomb pseudopotential of $K_3C_{60}$. Therefore, the electronic correlation should also be considered to fully understand alkali-fulleride superconductors.

In conclusion, we have comprehensively studied the band structure of alkali-doped $C_{60}$ ultrathin films. In the superconducting $K_3C_{60}$ films, we directly visualize the crucial band structure and band renormalization effects due to strong electron-phonon coupling. The analysis of ARPES spectra estimates the dimensionless electron-phonon coupling parameter to be about 1.2, evidencing the strong-coupling superconductivity of $K_3C_{60}$, supported by the large reduced superconducting gap. Our results also reveal signatures of electron correlation, suggesting unusual interplay between the electron-phonon interaction and electronic correlation in the electronic structure and superconductivity of $K_3C_{60}$ [13].

**Acknowledgements**

This work is funded by the National Key R&D program of China (Grants No. 2022YFA1403100, No. 2022YFA1403200, and No. 2017YFA0304600), the National Natural Science Foundation of China (Grants No. 12275148, and No. 11774190), and EPSRC Platform Grant (Grant No. EP/M020517/1). L. X. Y. acknowledges the support from Tsinghua University Initiative Scientific Research Program.

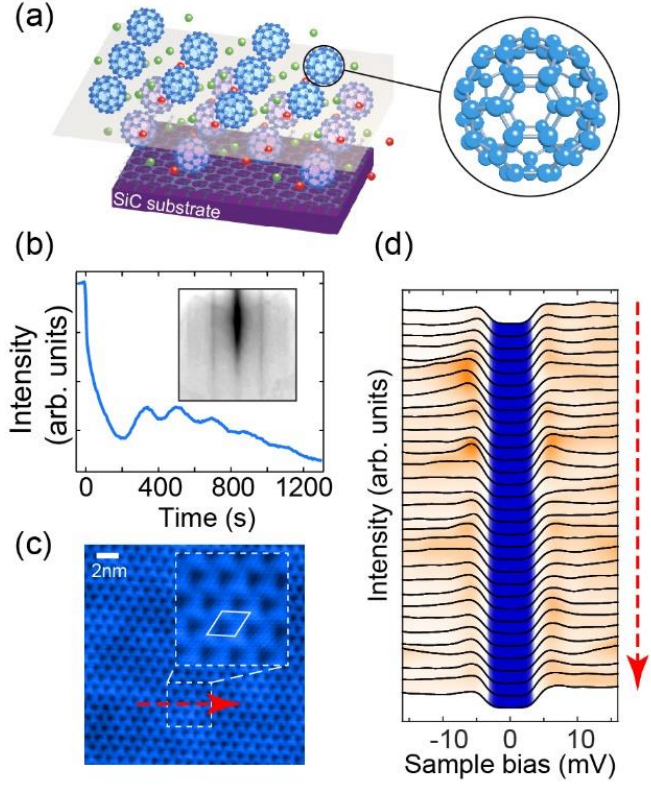

FIG. 1. (a) Schematic illustration of the crystal structure of bilayer $K_3C_{60}$ grown on an epitaxial bilayer graphene that was prepared by graphitizing the SiC substrate. The red and green spheres are the K atoms occupying octahedral and tetrahedral interstitial sites between $C_{60}$ molecules. The zoom-in plot shows a $C_{60}$ molecule with a diameter of about 10 Å. (b) The intensity of the specular spot in RHEED pattern as a function of sample growth time. The film thickness can be monitored by the oscillation in the RHEED curve. (c) STM surface topography of 5-layer $K_3C_{60}$ showing tri-lobe feature, acquired at sample bias $U = 1.5$ V and a constant tunneling current of $I = 30$ pA. (d) Line profile along the red dashed line in (c) showing the homogenous superconducting gap. Data were collected at 4.7 K.

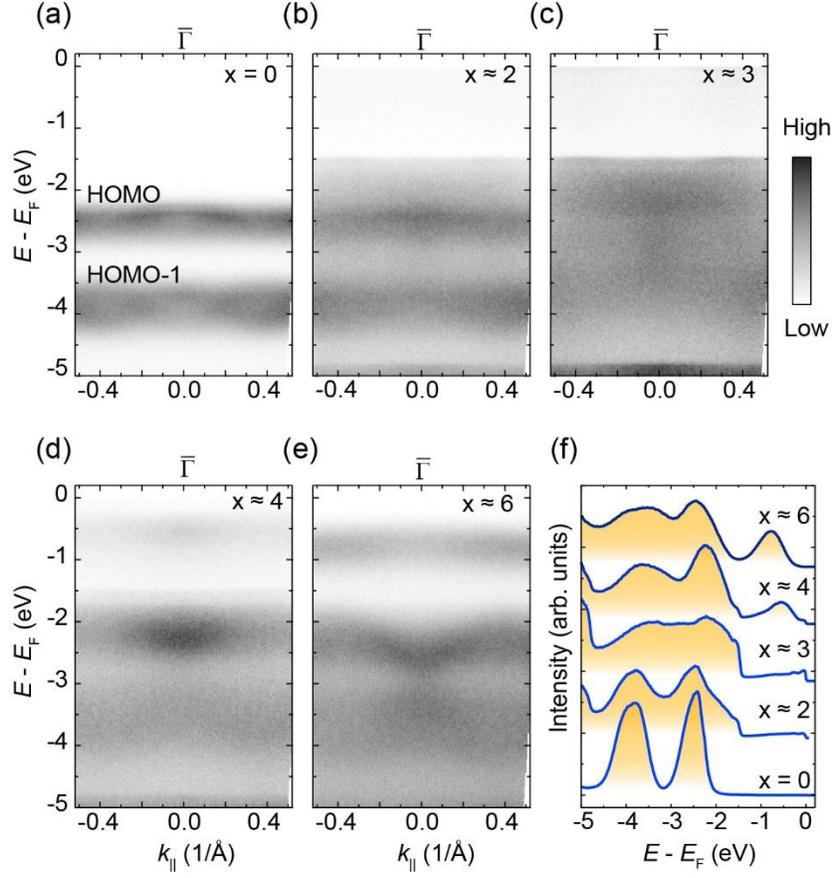

FIG. 2. (a) Band structure of 3ML pristine $C_{60}$ film grown on bilayer graphene/SiC substrate. (b-e) Evolution of the band structure of $K_xC_{60}$ film with K doping. (f) Integrated EDCs of $K_xC_{60}$ films in (a-e). Data were collected using helium lamp ($h\nu = 21.2$ eV) at 13 K.

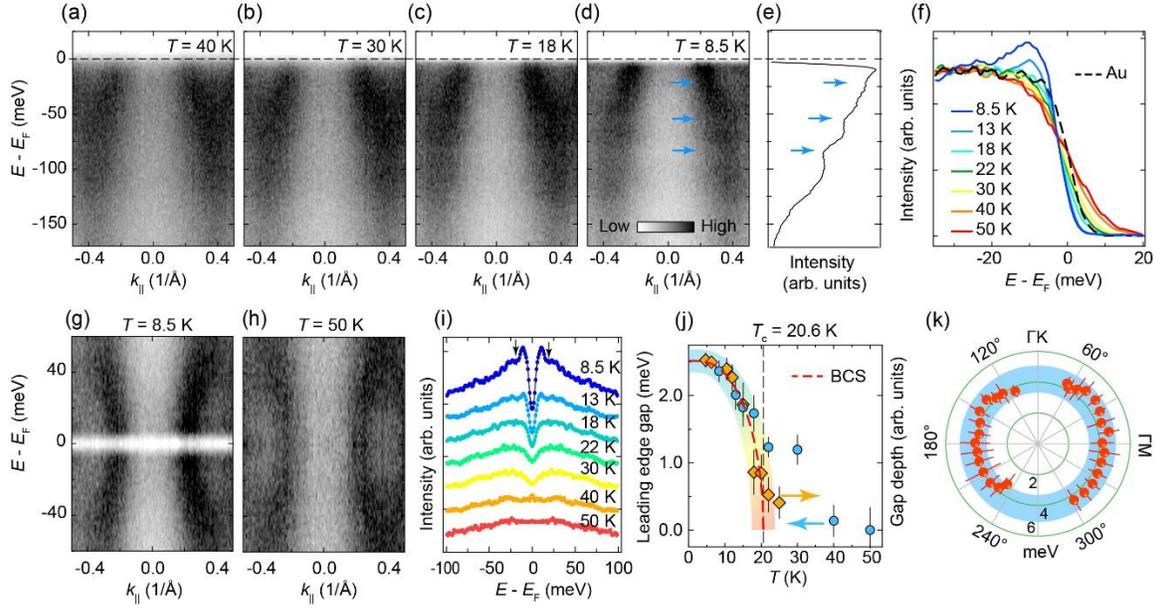

FIG. 3. (a-d) Band dispersion of 5ML $K_3C_{60}$ film along $\overline{\Gamma M}$ near $E_F$ collected at selected temperatures. (e) Integrated energy distribution curves (EDCs) at 8.5 K showing the reduction of spectral weight near binding energy 18, 54, and 85 meV. (f) Comparison between the EDCs of $K_3C_{60}$ at Fermi momentum ($k_F$) and polycrystalline gold showing the superconducting gap. (g), (h) Symmetrized ARPES spectra at 8.5 K and 50 K, respectively. (i) Symmetrized EDCs at $k_F$ acquired at selected temperatures. The black dashed line is the fit to the Dynes function. The black arrows indicate the dip in the EDC. (j) Temperature dependent leading-edge gap from ARPES measurements (blue circles) and gap depth from STM measurements (orange diamonds). The red dashed line is the fit to BCS model. (k) Angular distribution of superconducting gap extracted by fitting the symmetrized EDCs to the Dynes function.

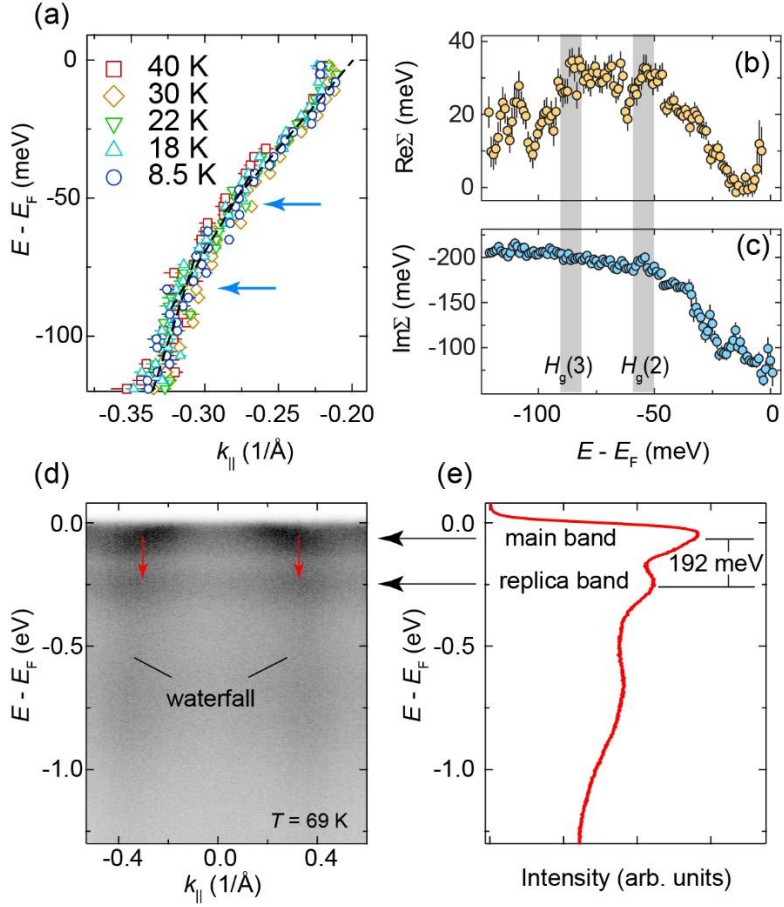

FIG. 4. (a) Comparison of the band dispersions at selected temperatures extracted from MDC fitting. The black dashed lines are the guides of eyes for the energy kinks in the band dispersion. (b), (c) Real and imaginary parts of the electron self-energy, respectively. (d) ARPES spectra in a large energy scale showing the band replication and waterfall-like band dispersion. (e) Integrated EDC obtained from (d).

supercondcting gap, (iv) temperature-dependent scanning tunnelling spectroscopy measurement, (v) details about the fitting of the band dispersion, (vi) extraction of electron-phonon coupling constant, (vii) phase separation at slight K doping.